\documentclass[a4paper,twoside,10pt]{article}
\usepackage{graphicx,publaob}

\def\papertype{Contributed paper}
\begin{document}

% Title
\title{ON THE COEXISTENCE OF PHASES IN A LENNARD JONES FLUID:FIRST RESULTS}

% Authors
\authors{V. \v CELEBONOVI\'C}

% Addresses and e-mails
\address{Institute of Physics,Pregrevica 118, 11080 Zemun-Belgrade, Serbia}
\Email{vladan}{phy.bg.ac}{yu}

% Running titles
\markboth{ON THE COEXISTENCE OF PHASES IN A L-J FLUID}{V. \v CELEBONOVI\'C }

% Abstract
\abstract{The aim of this paper is to investigate the conditions for the coexistence of phases in a Lennard Jones fluid.The calculation has been performed within the virial developement method,and as a result,a simple approximate relation has been obtained between the number densities of two coexisting phases  and the interparticle interaction potentials in them. The results of this work could have implications for modelling of giant planet interiors.This has become important,due to the discovery of more than 300 extrasolar planets.
}

% Section and subsection
\section{INTRODUCTION}

In the last 5-6 years of the $XX$ century,interest in planetology has drastically risen. This is due partially to results obtained within our planetary system,but even more so to discoveries of extrasolar planets. Until the end of November 2008., according to data at http://exoplanet.eu , $329$ exoplanets have been detected and 269 of them have masses $M\leq5 M_{J}$ where $M_{J}$ is the mass of Jupiter. This testifies about the interest of modelling the internal structure of the giant planets. For a recent study see,for example,(Vorberger et al.,2007 or Nettlemann et al.,2008).

The aim of this contribution is to present preliminary results on the conditions under which two phases in a Lennard-Jones fluid can be in equilibrium. Phases are defined as regions of the parameter space within which properties of a material are uniform.The condition for their equilibrium is the equality of pressures,temperatures and chemical potentials. In this contribution only the equality of pressures and temperatures will be considered; accordingly the results will be only preliminary. The equality of  chemical potentials will be included in future work.%In principle,phases can interact mutually,and contain an arbitrary number of components. It will be assumed in the following that the phases of the system are mutually non interacting,and that each of them contains only one component.
\section{METHOD}

The virial development of the equation of state is a method in which the equation of state $(EOS)$ of a fluid can be expressed as a power series in the density,and the coefficients take into account the interactions present in the system,in which an increasing number of particles takes part (e.g. Reichl 1988). 
\footnotetext[1]{Presented at the $15^{th}$ National Conf.of Astronomers of Serbia,Belgrade,October $2$-$5.$,$2008$.The proceedings will appear in Proc.Astron.Obs.Belgrade.Disregard the volume and page numbers.}

The mathematical form of the $EOS$ of a fluid within the virial development is
\begin{equation}
	\frac{p v}{k_{B} T}= \sum_{l=1}^{\infty}a_{l}(T)(\frac{\lambda^{3}}{v})^{l-1} 
\end{equation}
All the symbols on the left side of Eq.(1) have their standard meanings,while on the right hand side,$a_{l}$ are the so called virial coefficients,$\lambda$ is the thermal wavelength and $v$ is the inverse number density of the system $v=V/N$. The thermal wavelength is given by (for example Reichl,1988) 
\begin{equation}
	\lambda=(\frac{2\pi\hbar^{2}}{m k_{B} T})^{1/2}
\end{equation}
where $\hbar$ is Planck's constant and $m$ the particle mass.
The $LJ$ model potential has the form
\begin{equation}
	u(r)=4\epsilon\left[(\frac{\sigma}{r})^{12}-(\frac{\sigma}{r})^{6}\right] 
\end{equation} 
The symbol $\epsilon$ denotes the depth of the potential,while $\sigma$ is the diameter of the molecular "hard core". 
The first coefficient in Eq.(1) is $a_{l}=1$ ,while the second one is given by
\begin{equation}
	a_{2}(T)=-(2/3) \pi N_{A} \beta \int_{0}^{\infty}\exp(-\beta u)(\frac{\partial u}{\partial r}) r^{3} dr
\end{equation}
(Maitland,G,C.,Rigby,M.,Smith.B.E.and Wakeham,W.A.,1987). In this and other expressions $\beta=1/k_{B} T$ ,$T$ is the temperature and $k_{B}$ is the Boltzmann constant.It can be shown that the second virial coefficient for the Lennard-Jones potential is given by 
\begin{equation}
	a_{2}(T^{*}) = b_{0}\sum_{j=0}^{\infty}\gamma_{j} (1/T^{*})^{(2j+1)/4} 
\end{equation}
where
\begin{equation}
	\gamma_{j}=\frac{-2^{(j+1/2)}}{4j!} \Gamma(\frac{2j-1}{4}) 
\end{equation}

\begin{equation}
	T^{*}=\frac{1}{\beta\epsilon} 
\end{equation}
and $b_{0}=(2 \pi/3) N_{A} \sigma^{3}$,where $N_{A}$ is Avogadro's number.
The chemical potential of a fluid is given by (Hill,1987):
\begin{equation}
	\frac{\mu}{k_{B}T}=\ln(n \lambda^{3})+\frac{n}{k_{B}T} \int_{0}^{1}d\gamma\int_{0}^{\infty}dr 4\pi r^{2} u(r) g(r)
\end{equation}
where $n$ is the particle number density and $g(r)$ is the radial distribution function. The problem with the calculation of $\mu$ is the determination of $g(r)$,which is a complicated task in statistical mechanics.In the future,it will be attempted to insert the form of $g(r)$ proposed in (Morsali et.al.,2005) into Eq.(8).  
\section{THE CALCULATION}  

%\subsection{The definition of the problem}

%\subsection{The calculation} 

In order to render the $EOS$ within the virial development physically applicable, the power series in Eq.(1) has to be convergent. This means that the results are applicable only under the condition

\begin{equation}
	\frac{\lambda^{3}}{v}=n\lambda^{3}\prec1
\end{equation}

Inserting Eq.(2) into Eq.(9),it follows that the virial development is applicable under the conditions 

\begin{equation}
	n\prec\left(\frac{m }{2\pi}\right)^{3/2} \left[\frac{(k_{B}T)^{1/2}}{\hbar}\right]^3 
\end{equation}

Denote the two mutually non-interacting phases which make up the system by  $"1"$ and $"2"$. Applying Eq.(1),gives
\begin{equation}
	p_{1}= n_{1} k_{B} T_{1}\sum_{l=1}^{\infty}a_{l}(T)(\frac{\lambda_{1}^{3}}{v_{1}})^{l-1} 
\end{equation}

and 
\begin{equation}
	p_{2}= n_{2} k_{B} T_{2}\sum_{r=1}^{\infty}c_{r}(T)(\frac{\lambda_{2}^{3}}{v_{2}})^{r-1} 
\end{equation}
Inserting the conditions of the equality of pressures and temperatures needed for the coexistence in equilibrium of the two phases ,one gets the following expression for the ratio of densities in them
\begin{equation}
	\frac{n_{1}}{n_{2}}=\frac{\sum_{r=1}^{\infty}c_{r}(T)(n_{2}\lambda_{2}^{3})^{r-1}}{\sum_{l=1}^{\infty}a_{l}(T)(n_{1}\lambda_{1}^{3})^{l-1}}
\end{equation}
Taking into account that the equality of the temperatures implies the equality of the thermal wavelengths, this expression can be transformed into the following form
\begin{equation}
	\sum_{l=1}^{\infty}a_{l}(T)n_{1}^{l} \lambda^{l-1}=	\sum_{r=1}^{\infty}c_{r}(T)n_{2}^{r} \lambda^{r-1} 
\end{equation}
Limiting the sums to the first two terms,it follows that
\begin{equation}
	a_{1}n_{1}+a_{2}n_{1}^{2} \lambda=c_{1}n_{2}+c_{2}n_{2}^{2}\lambda
\end{equation}
Taking into account that the first virial coefficient is 1,and introducing $n_{1}-n_{2}=x$, the last expression can be solved to give
\begin{equation}
	x=(1/2\lambda a_{2})\left[-1-2\lambda a_{2}n_{2}+\sqrt{1+4 a_{2}n_{2}\lambda(1+c_{2}n_{2}\lambda)}\right]
\end{equation}
This result is mathematically simple,but physically interesting. It gives the difference between the number densities of  two phases  coexisting in equilibrium,expressed in terms of the density of one of the phases ,the thermal wavelength and the second virial coefficient in both of them. Expression (16) can further be transformed to give finally
\begin{equation}
	n_{1}=-\frac{1}{2\lambda a_{2}}\left[1-\sqrt{1+4 a_{2}n_{2}\lambda (1+c_{2} n_{2}\lambda)}\right]
\end{equation}
which is positive for $4 a_{2} n_{2} (1+c_{2} n_{2}\lambda)\succ0$. 

This result is even more physically interesting. It represents a link ( of second order,because only the second virial coefficients are taken into account) of the number densities in two phases  of a LJ fluid,expressed as a function of the temperature and the interaction potentials in both of them. This result is general,in the sense that virial coefficients for {\bf any} potential can be inserted in it. 

Turning to the case of a $L-J$ fluid,the values of the first few coefficients $\gamma_{j}$ are: $\gamma_{0}=1.733$,$\gamma_{1}=-2.564$,$\gamma_{2}=-3.466$. This means that the explicite expression for the second virial coefficient is
\begin{equation}
	a_{2}=\frac{2\pi}{3} N_{A}\sigma^{3} \left[1.733(\frac{\epsilon}{k_{B}T})^{1/4}-2.564(\frac{\epsilon}{k_{B}T})^{3/4}-3.466(\frac{\epsilon}{k_{B}T})^{5/4}+...\right]
\end{equation}
Inserting Eq.(18) into Eq.(17),one would obtain an expression "linking" the number densities of the two phases  coexisting in equilibrium with the parameters of interparticle potentials in them.
How could this result be applied in studies of the interiors of the Jovian planets?
Within any celestial body, the particle number density increases with increasing depth. Therefore, the parameters of the $L-J$ potential would also be density dependent,which implies that the virial coefficients would also be depth (an density) dependent.This means that the possibility of  phase coexistence would be density dependent.A closely related problem is the density dependence of the phase transition pressure in a system with a $L-J$ potential. These and other issues related to problems of the behaviour of a $L-J$ fluid under high density will be studied in future work.  
%\vfil\newpage

% Table
%\begin{table}
%\begin{center}
%\begin{tabular}{|c|c c|}
%\hline
%$t$    &$\theta (^o)$ & $\rho ('')$ \\
%\hline
%2002.0 &    174.0     &    0.259    \\
%2003.0 &    174.7     &    0.262    \\
%2004.0 &    175.3     &    0.265    \\
%2005.0 &    175.9     &    0.268    \\
%2006.0 &    176.6     &    0.270    \\
%\hline
%\end{tabular}
%\caption{Calculated positions for the next five years}
%\end{center}
%\end{table}

% Figure (in PS or EPS format)
%\begin{figure}
%\includegraphics[width=12cm,height=8cm]{fig1.eps}
%\caption{The observed $Fe K_\alpha$ line and its fit}
%\end{figure}

% Equation
%\begin{equation}
%G(x)=\int_0^\infty e^{-{{x^2}\over{2}}} dx
%\end{equation}
\section{Acknowledgement}

This contribution has been prepared within the research project 141007 financed by the Ministry of Science,Technology and Development of Serbia. I am grateful to the referee for helpful comments.
% References
\references

Vorberger,J.,Tamblin,I.,Militzer,B.and Bonev,S.A.: 2007,\journal{Phys.Rev.},{\bf B75},024206.

Nettlemann,N.,Holst,B.,Kietzmann,A.,French,M.,Redmer,R.and Blaschke,D.: 2008,

\journal{Astrophys.J.},{\bf 683},1217.

Reichl,L.E.:1988,\journal{A Modern Course in Statistical Physics},Edward Arnold (Publishers) Ltd.,

Great Britain. 

Maitland,G,C.,Rigby,M.,Smith.B.E.and Wakeham,W.A.:1987,\journal{Intermolecular Forces:}

\journal{Their Origin and Determination},Oxford Science Publications.

Hill,T.L.:1987,\journal{Statistical Mechanics:Principles and Selected Applications},Dover 

Publications Inc.,New York.

Morsali,A.,Goharshadi,E.K.,Ali Mansoori,G.and Abbaspour,M.:
2005,\journal{Chemical Physics},

\vol{310},11. 
%Hewitt, A., Burbrdge, G. : 1989, \journal{Astrophys. J. Suppl. Series}, \vol{75}, 297.

%Mediavilla, E., Insertis, F. M. : 1989, \journal{Astron. Astrophys.} \vol{214}, 79. 

%Netzer, H. : 1990, \journal{Active Galactic Nuclei, eds. R. D. Blandford, H. Netzer \& L. Woltjer}, Saas-Fee Advanced Course 20, Berlin: Springer -- Verlag.  

%Osterbrock, D. E. : 1989, \journal{Astrophysics of Gaseous Nebulae and Active Galactic Nuclei}, Mill Valley, California. 
\endreferences

\end{document}